\documentclass[pre, twocolumn, amsmath, amssymb, superscriptaddress]{revtex4}

\usepackage{graphicx}
\usepackage{dcolumn}
\usepackage{bm}
\usepackage{braket}
\usepackage{xcolor}

\usepackage{amsmath}

\newcommand{\beq}{\begin{equation}}
\newcommand{\eeq}{\end{equation}}

\newcommand{\uv}[1]{{ \boldsymbol{ #1}}}

\newcommand{\bpm}{\begin{pmatrix}}
\newcommand{\epm}{\end{pmatrix}}

\newcommand{\rv}[1]{{\bf r}}
\newcommand{\qv}[1]{{\bf q}}

\begin{document}

\title{Rosette formations as symmetry-breaking events: theory and
 experiment}

\author{Mattia Miotto \footnote{\label{coau} These authors contributed equally to the present work.}\footnote{\label{cocorr} Corresponding authors: mattia.miotto@roma1.infn.it; matteo.paoluzzi@uniroma1.it} }

\affiliation{Center for Life Nano \& Neuro Science, Istituto Italiano di Tecnologia, Viale Regina Elena 291,  00161, Rome, Italy}

\author{Giorgio Gosti\footref{coau}}
\affiliation{Institute of Heritage Science, National Research Council (CNR-ISPC), Via Salaria KM 29300, 00015 Monterotondo, Italy}
\affiliation{Center for Life Nano \& Neuro Science, Istituto Italiano di Tecnologia, Viale Regina Elena 291,  00161, Rome, Italy}

\author{Maria Rosito}
\affiliation{Center for Life Nano \& Neuro Science, Istituto Italiano di Tecnologia, Viale Regina Elena 291,  00161, Rome, Italy}
\affiliation{Department of Life Sciences, Health and Health Professions, Link Campus University, Via del Casale di San Pio V, 44 00165, Rome, Italy}

\author{Michela Dell'Omo}
\affiliation{Department of Basic and Applied Sciences for Engineering, Sapienza University, Rome, Italy}
\affiliation{Center for Life Nano \& Neuro Science, Istituto Italiano di Tecnologia, Viale Regina Elena 291,  00161, Rome, Italy}

\author{Viola Folli}
\affiliation{Center for Life Nano \& Neuro Science, Istituto Italiano di Tecnologia, Viale Regina Elena 291,  00161, Rome, Italy}
\affiliation{D-tails s.r.l.,Rome,Italy}

\author{Valeria de Turris}
\affiliation{Center for Life Nano \& Neuro Science, Istituto Italiano di Tecnologia, Viale Regina Elena 291,  00161, Rome, Italy}

\author{Giancarlo Ruocco}
\affiliation{Department of Physics, Sapienza University, Piazzale Aldo Moro 5, 00185, Rome, Italy}
\affiliation{Center for Life Nano \& Neuro Science, Istituto Italiano di Tecnologia, Viale Regina Elena 291,  00161, Rome, Italy}

\author{Alessandro Rosa}
\affiliation{Department of Biology and Biotechnologies ``Charles Darwin'', Sapienza University,  Piazzale Aldo Moro 5, 00185, Rome, Italy}
\affiliation{Center for Life Nano \& Neuro Science, Istituto Italiano di Tecnologia, Viale Regina Elena 291,  00161, Rome, Italy}

\author{Matteo Paoluzzi\footref{cocorr}}
\affiliation{Department of Physics, Sapienza University, Piazzale Aldo Moro 5, 00185, Rome, Italy}

\begin{abstract}
Multicellular rosettes are ubiquitous structures in biological systems, observed in contexts ranging from morphogenesis to wound healing and cancer progression. While some molecular insights have been gained to explain the presence of these assemblies of cells around a common center, what are the tunable, global features that favor/hinder their formation is still largely unknown.
Here, we made use of a Voronoi dynamical model to investigate the ingredients driving the emergence of rosettes characterized by different degrees of stability and organization. 
We found that (i) breaking the local spatial symmetry of the system, i.e., introducing curvature-inducing defects, allows for the formation of rosette-like structures (ii) whose probability of formation depends on the characteristics of the cellular layer. In particular, a trade-off between tissue fluidity and single-cell deformability dictates the assembly of transient rosettes, which are strongly stabilized in the presence of cell self-alignment interactions.  

To test our model predictions, we performed fluorescence microscopy experiments on rosette-forming neural cell populations derived from induced pluripotent stem cells, finding significant agreement with the model predictions. Overall, our work may set the stage to gain an unifying understanding of the plethora of biophysical mechanisms involving the occurrence of rosette-like structures in physiology and their altered formation in pathology.
\end{abstract}

\maketitle

\section{Introduction}

Rosettes are polarized, transient epithelial structures, observed as intermediate states in several biological processes as different as morphogenesis, development, wound healing, and cancer~\cite{Harding2014, Tetley2019, Gredler2023, Moran2023}.
From a geometrical point of view, these structures can be identified in 2D sections of cellular layers as 
assemblies composed of five or more cells that interface at a central point~\cite{Leng2021}. 

The mechanism through which such a configuration can be reached is proposed to be a topological rearrangement following either the disappearance of a cell (T2 transition) due to extrusion, ablation, or cell apoptosis~\cite{Harding2014} or intercalation events where multiple T1 transitions lead to a 5 or more-fold vertex ~\cite{Blankenship2006, Fletcher2014}. The latter preserves the number of cells in the population.

Theoretical and numerical investigations so far have focused on 
studying the effect the presence of rosettes has on the tissue dynamics. Using modified vertex models, it has been shown that rosettes have an impact on tissue fluidity~\cite{Yan_2019} and reorganization~\cite{Trichas2012, Siang2018}. 

Indeed, transient rosettes are known to either resolve, producing tissue reorganizations~\cite{Siang2018} or stabilize in lumen-centered polarized rosettes~\cite{Harding2014}, a critical process for organ development, including those of the kidneys, mammary glands, and neural tube \cite{MartinBelmonte2008, Bryant2010}.  Disregulation in such a process often results in severe developmental defects~\cite{Avagliano2018}. 
Despite their centrality, the environmental conditions, i.e., the extracellular signals and cellular events, driving rosette formation are yet to be fully determined~\cite{Miotto2023}.


\begin{figure*}[t]
\centering
\includegraphics[width = \textwidth]{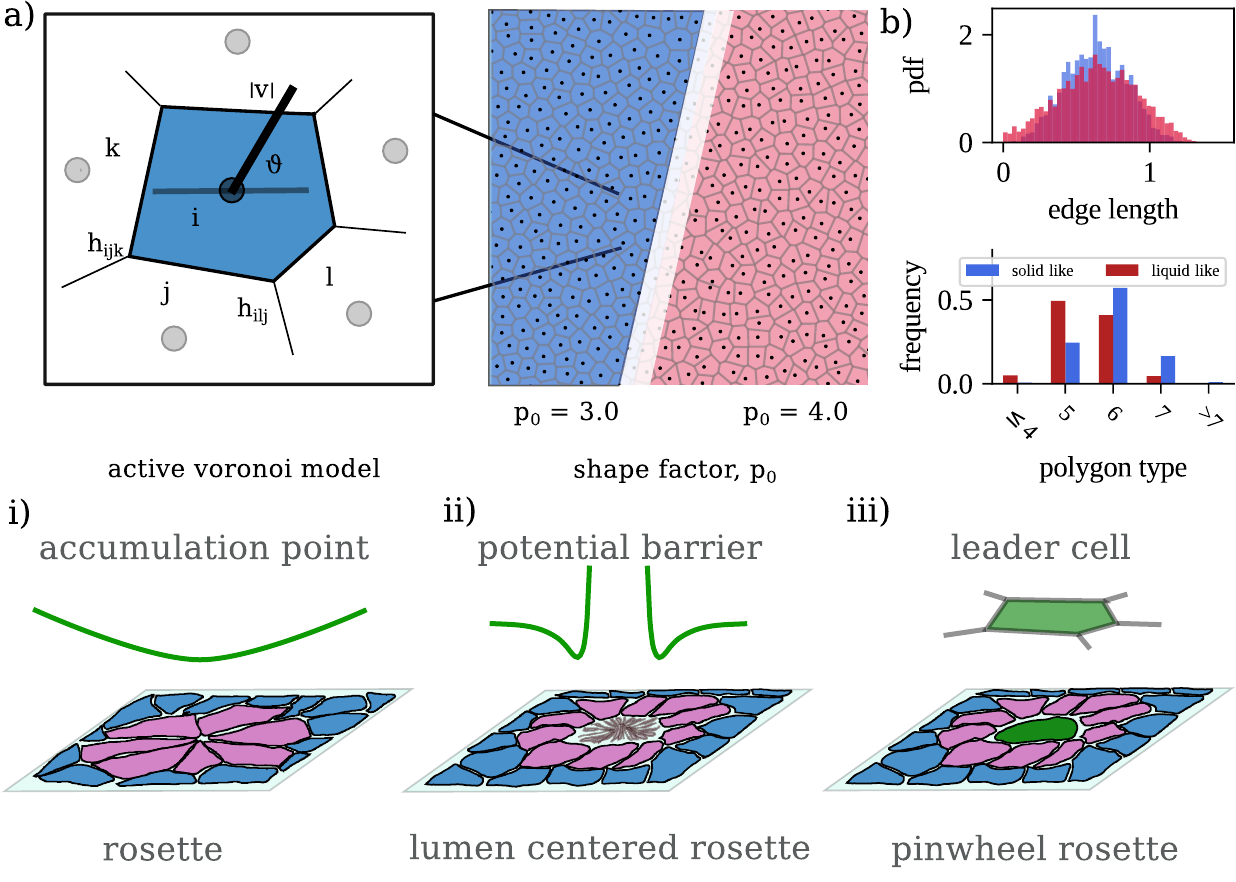}
\caption{\textbf{Modeling rosette formation.} \textbf{a)} Examples of stationary configurations in the solid and fluid-like regimes are shown with cells colored in blue and red, respectively. \textbf{b)}  (top) Probability density distribution (pdf) of the cells' edge lengths for the configurations shown in panel a). (bottom) Frequency of occurrence of cells as a function of their edge number. \textbf{c)} Sketches of three possible rosette conformations and the corresponding used model. (i) Transient rosette configurations, with 5 or more cells sharing a vertex, are simulated by breaking spatial symmetry via a harmonic potential acting on a given point. (ii) Rosettes with a central lumen are obtained via a Lennard-Jones potential energy originating in the center of the lumen, while (iii) pinwheel rosettes are simulated by inserting in the simulation a leader cell having a different target area.       
}
\label{fig:1}
\end{figure*}



Here, for the first time, we make use of the Active Voronoi model to search for the minimal ingredients needed for the formation of rosette-like structures in 2D layers of epithelial cells. 
In contrast to vertex models, where topological
transitions have to be incorporated by hand, the Voronoi-based approach deals with transitions automatically as cell centers instead of cell vertices, making this coarse-grain model adapt to study rosettes dynamics without the need for parameter-richer models like those based on phase-fields~\cite{Wenzel2019}. Voronoi models are very effective as a coarse-grained description of confluent monolayers \cite{Park15,Bi2016,Malinverno17,giavazzi2018flocking,PhysRevMaterials.2.045602}. However, additional ingredients are required for triggering rosette formation in Voronoi models.

We isolate two key ingredients for observing such structures in monolayers through Voronoi models. The first ingredient is the breaking of the space translational invariance that generates topological defects. Those topological defects are candidates for the formation of rosette-like structures. However, without another dynamic ingredient that stabilizes those defects, rosettes are hindered by cell motility and deformability. We observe that self-alignment interactions, producing cell displacements in the same direction along which cells are elongated, are good candidates as minimal mechanisms for stabilizing those defects.
We explored three paradigmatic cases in which the formation of rosettes is induced by breaking the spatial translational invariance of the system via the introduction of a privileged point: (i) the case of harmonic field, (ii) the case of a Lennard-Jones (LJ) like potential that contains both, repulsion and attraction, and (iii) the case of a leader cell.  
In particular, harmonic and LJ-like potentials localized at one point of the space produce the formation of transient and lumen-centered rosettes, respectively. In the case of a leader cell,
we considered a different intrinsic size with respect to the rest of the population,  mimicking the behavior of pinwheel rosettes~\cite{Harding2014}.
For each of the three symmetry-breaking scenarios, we probed the effects of cell self-velocity, shape index, self-alignment interaction, and persistence length, which are well-established control parameters in cell tissue models, to isolate the minimal ingredients needed for observing stable rosette-like structures.
We found that in the absence of self-alignment interaction, (1) transient rosette formation is permitted in confluent layers of deformable, slowly moving cells. The probability of observing rosettes in motile, rigid cells scales with the intensity of the external potential/strength of the attractive force toward a specific point. 
(2) Alignment interaction both increases the probability of transient rosette formation and the size of lumen-centered ones. In particular, it allows cells to be orderly disposed in multiple layers around the lumen.  Finally, to test our model predictions, we performed experiments on iPSC-derived neural cells that are able to form rosettes after approximately eleven days of differentiation. Performing extensive fluorescent microscopy time lapses, we found (3) that neural rosettes are observed when the iPSCs populate the region of the parameter space our model predicts to be compatible for their formation.

\begin{figure*}[t]
\centering
\includegraphics[width = \textwidth]{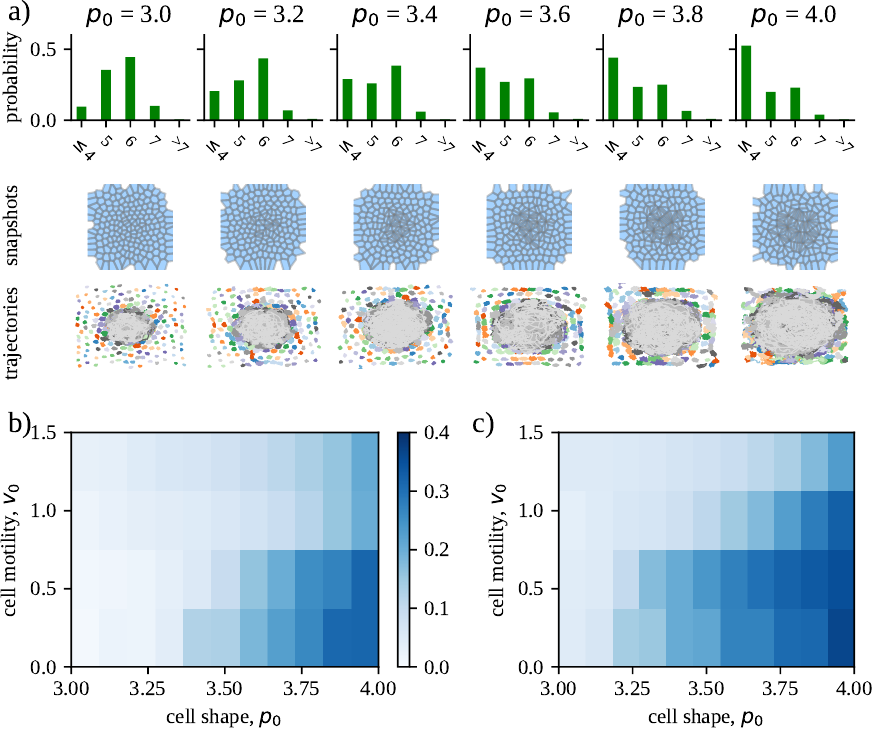}
\caption{\textbf{Trade-off between fluidity and stiffness} \textbf{a)}  From top to bottom, frequency of occurrence of cell as a function of their edges number, snapshot of representative stationary configurations, and trajectories upon varying the cell shape factor for a population of cells having self-propulsion velocity, $v0=0.5$ and persistence time, $\tau= 50$ in presence of an external harmonic potential of intensity, $k=0.2$. \textbf{b)} Probability of observing rosette as a function of cell shape and cell motility for a population of cells subject to an external harmonic potential of intensity, $k=0.2$. Colors range from white to blue as the probability increases.  \textbf{c)} Same as in panel b), but in the presence of a harmonic potential of intensity, $k=0.3$. 
}
\label{fig:2}
\end{figure*}

\begin{figure}[t]
\centering
\includegraphics[width = \columnwidth]{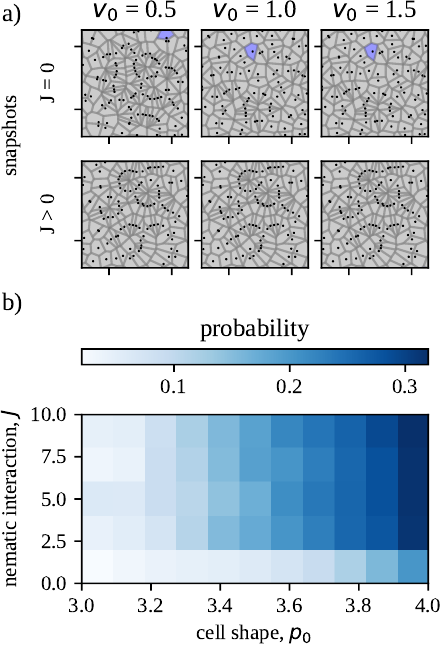}
\caption{\textbf{Alignment interaction favors rosette formation.} \textbf{a)} Representative stationary configurations for populations of cells having different motility with ($J>0)$ and without ($J=0$) self-alignment interaction in the presence of an external harmonic potential of intensity $k=0.2$. Observed rosettes are highlighted.
\textbf{b)}  Probability of observing rosette as a function of cell shape and self-alignment interaction strength for a population of cells subject to an external harmonic potential of intensity, $k=0.2$. Colors range from white to blue as the probability increases. 
}
\label{fig:3}
\end{figure}

\begin{figure*}[t]
\centering
\includegraphics[width = \textwidth]{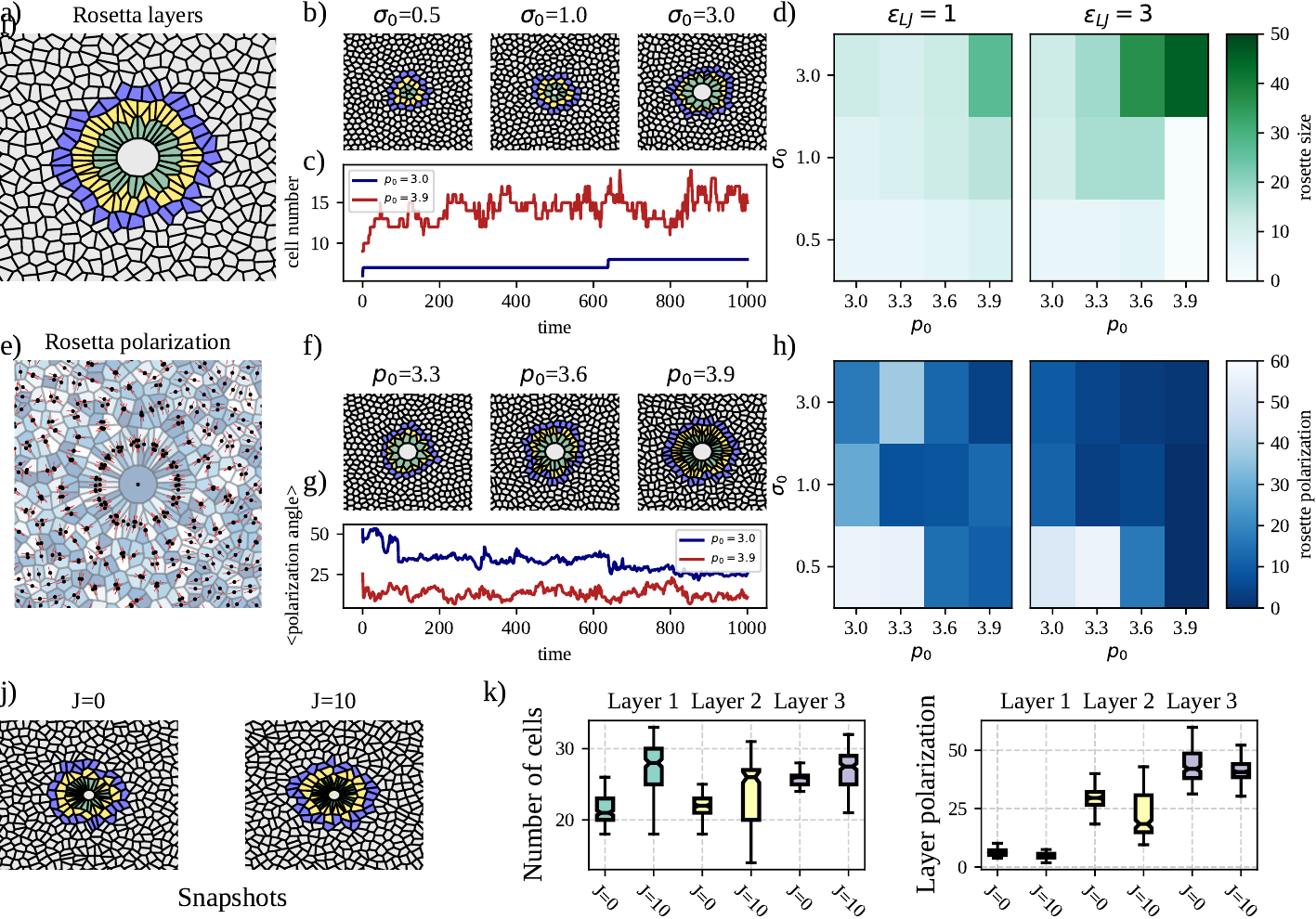}
\caption{\textbf{Effect of model parameters on lumen-centered rosettes.} \textbf{a)} Snapshot of a stationary configuration obtained via the Active Voronoi model in the presence of a Lennard-Jones potential. Cells forming the first, second, and third layers around the central lumen are colored in green, gold, and blue, respectively.   \textbf{b)} Snapshots obtained for different values of the LJ potential effective radius, $\sigma_0$. \textbf{c)}  Number of cells in the first cellular layers as a function of time for different values of the cell target shape factor, $p_0$. \textbf{d)} Stationary number of cells in the first cellular layers as a function of the cell target shape factor, $p_0$, and LJ effective radius, $\sigma_0$, for two values of the Lennard-Jones potential strength, $\epsilon_{LJ}$. 
\textbf{e)} Same as in panel a) but with cells colored according to their elongation-alignment with respect to the lumen center. 
\textbf{f)} Snapshots obtained for different values of the cell target shape factor, $p_0$, in the presence of a Lennard-Jones potential of effective radius, $\sigma_0=3.0$. 
\textbf{g)}  Average polarization of cells in the first cellular layers as a function of time for different values of the cell target shape factor, $p_0$.
\textbf{h)} Same as in panel d) but for the rosette polarization index.
\textbf{j)} Same as in panel a) but for different values of the cell self-alignment.
\textbf{k)} Boxplot representation of the distribution of 
the number of cells composing the rosette layers and their polarization indexes in the absence and presence of self-alignment interaction.
} 
\label{fig:4}
\end{figure*}

\begin{figure*}[]
\centering
\includegraphics[width = 0.8\textwidth]{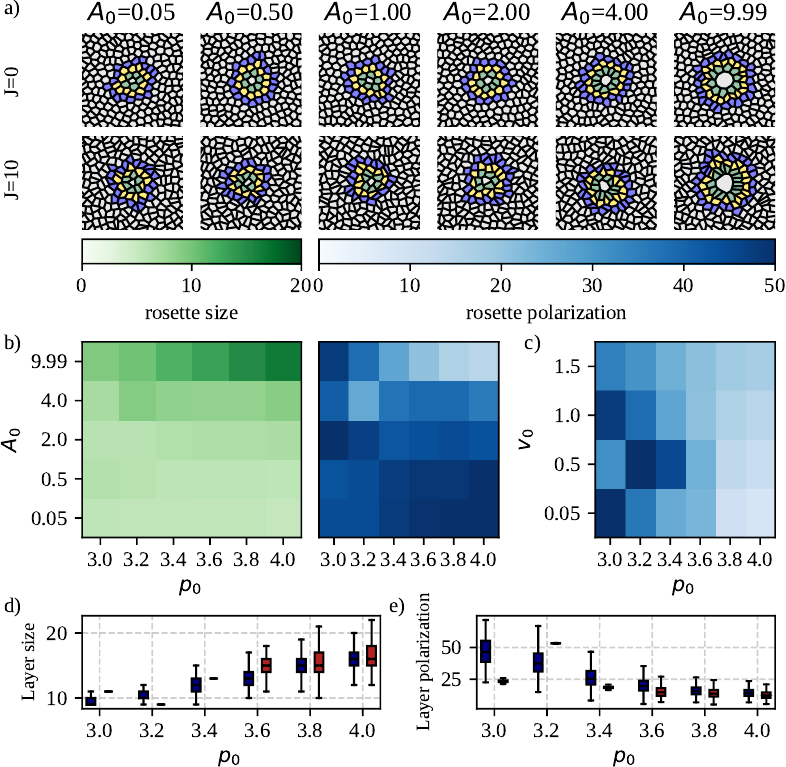}
\caption{\textbf{Comparison between lumen and leader cell-centered rosettes.} \textbf{a)} Snapshots of stationary configurations obtained via the Active Voronoi model in the presence of a leader cell of target area $A_0$. The top row displays snapshots for cells having a target shape factor $p_0=3.2$, while the bottom row corresponds to cells with $p_0=3.8$. Cells forming the first, second, and third layers around the central lumen are colored in green, gold, and blue, respectively. \textbf{b)}  Stationary number of cells in the first cellular layers (green shaded map) and polarization index (blue shaded map) as a function of the cell target shape factor, $p_0$, and leader cell target area, $A_0$.  
\textbf{c)}  Stationary polarization index of the first cellular layers as a function of the cell target shape factor, $p_0$, and single cell velocity, $v_0$, for a population of cells in the presence of a leader cell with target area, $A_0 = 10$.    \textbf{d)} Boxplot representation of the distribution of 
the number of cells composing the rosette first layers in the absence and presence of self-alignment interaction for different values of the cells' target shape factor, $p_0$.
\textbf{e)} Same as in panel d) but for the polarization index of the cells in the rosette first layer.
} 
\label{fig:5}
\end{figure*}

\section{Results}

\subsection{Rosette formation in the Voronoi Model} 
We consider a population of $N$ epithelial cells confined in a square box of side $L\!=\!\sqrt{N}$ with periodic boundary conditions. We indicate with ${\boldsymbol{r}}^i$ the position of the $i-$th cell center (with $i\!=\!1,..,N$) whose equation of motion reads \cite{Bi2016}
\begin{equation}
    \label{eq:dotr}
    \dot{\boldsymbol{r}}^i = v_0^i \hat{e}_i + \mu \uv{F}_i 
\end{equation}
where $\uv{F}_i$ is the total force acting on the cell, $\mu$ the mobility,  
and  $v_0^i$  the modulus of the self-propulsion velocity whose direction in 2D is given by the versor $\hat{e}_i$, parameterized by the angle $\theta_i$, as  $\hat{e}_i = (\cos\theta_i, \sin\theta_i)$. The modulus of the self-propulsion velocity is assumed to be the same for all cells,  $v_0^i=v_0$.

Cell centers undergo Active Brownian dynamics in which the orientation of the self-propulsion
$\theta_i$ follows a standard Langevin equation of the form $\dot{\theta}_i = \eta_i$, with  $\langle \eta_i \rangle = 0$ and $\langle\eta_i(t)\eta_j(s) \rangle = \frac{2}{\tau}\delta_{ij}\delta(t-s)$. 
The persistence time of the active motion, $\tau$,  is the inverse of the rotational diffusivity coefficient $D_r$.

The total mechanical force acting on each cell regulates cell fluctuations and is given by:
\begin{equation}
\label{eq:tot_force}
    \boldsymbol{F}_i = -\nabla_{\boldsymbol{r}_i}  E_{v}
\end{equation}
with  the term  $E_v \!=\! \frac{1}{2} \sum_i^N E_v^i$ being the geometrical energetic contribution given by the Vertex energy:
\begin{equation}
\label{eq:vertex-ene}
    E_v^i \!=\!  K_p (p_i -p_0)^2 + K_A (A_i -A_0)^2 \; .
\end{equation}
In particular,  $p_i$ and  $A_i$ represent the perimeter and area of cell $i$, respectively, while $K_p$ and $K_a$  are constants depending on the specific cell type.
The area modulus $K_a$ determines how resistant a cell is to deviations from its preferred area. It penalizes deviations of the actual area of the cell from its target area $A_0$. The latter reflects the size that cells tend to maintain due to internal forces such as pressure, growth constraints, or packing density. Similarly,  $K_p$ is the perimeter elasticity constant or perimeter modulus, which controls how resistant a cell is to changes in its perimeter with respect to its preferred perimeter $p_0$. Analogous to the preferred area $A_0$, the target perimeter reflects the cell’s preferred shape in terms of its perimeter and typically accounts for the contractility of the actomyosin network at the cell boundary.
 Without losing generality, one can set $K_p$, $K_a$, and  $A_0$ to one, and keep $p_0$ as a control parameter. 
In Figure \ref{fig:1}a, we depict the relevant features of the model.
A system subject only to vertex energy displays a phase transition between a solid and liquid-like behavior (see Figure~\ref{fig:1}b) for $p_0\sim 3.81$ (for details, see ~\cite{Bi2016}).  To quantify the propensity of the model to form rosettes, we measured the frequency of occurrence of cells with less than 5 edges, $P_r$, along the line of Trichas \textit{et al.}~\cite{Trichas2012}. As one can see from the bar plot in Figure~\ref{fig:1}b, the probability of observing rosettes is low as the system tends to organize in a mixture of pentagon and hexagon-like shapes, which minimize the Vertex energy in translational invariant space, i.e., flat ones.
To induce rosette formation, we modified the topological features of the system in different manners, obtaining transient, lumen-centered, and pinwheel rosettes depending on the adopted defect source (see Figure~\ref{fig:1}i-iii).

\subsection{Formation of rosettes requires a tradeoff between tissue fluidity and single cell stiffness}

To promote the formation of transient rosettes, we modified Eq.~\ref{eq:tot_force} by adding an external harmonic potential centered at a point C:
\begin{equation}
\label{eq:tot_force_harm}
    \uv{F}_i = -\nabla_{\uv{r}_i}  (E_{v} + E_h)
\end{equation}

with $E_h= \sum_i \frac{1}{2} k (\uv{r}_i - \uv{r}_c)^2$ and $\uv{r}_c= [L/2, L/2]$.

The external potential may mimic the effect of the forces exerted by actin fibers,  the attraction produced by the diffusion of specific molecules in the cellular micro-environment, and/or the curvature of the organoid/embryo surface ~\cite{FernandezGonzalez2011}.
Figure~\ref{fig:2} shows the results obtained simulating a population of 400 cells having self-propulsion velocity, $v0=0.5$, and persistence time, $\tau= 50$, in the presence of an external harmonic potential of intensity, $k=0.2$.
Upon varying the cells' target perimeter, i.e., their shape index,  the frequency of occurrence of cells having a number of edges lower than 5, $P_r$, increases with $p_0$ (see Figure~\ref{fig:2}a). Consequently, rosette-like structures start appearing in the stationary configurations. Looking at the single cell trajectories, one clearly sees that cells in the rosette-forming region are highly motile, indicating that the formed assemblies are transient ones. Note that introducing an external field, i.e., a curvature, results in a fluidification of the cell layer even for cells with low target shape factor~\cite{Sussman2020, DeMarzio2025}. To further investigate the effect of fluidity on rosette formation, we explored the behavior of the system upon changing cell self-propulsion velocity, $v_0$. In the absence of an external potential, increasing $v_0$ allows the tissue to undergo a solid to fluid transition~\cite{Bi2016}. Figure~\ref{fig:2}b displays $P_r$ as a function of cell shape and cell motility for a population of cells subject to an external harmonic potential of intensity, $k=0.2$. While the probability systematically increases with cell shape deformability ($p_0$), motility dampens the capability of the cells to form rosettes. Indeed, varying the strength of the harmonic potential shows that there is a tradeoff between cell motility and harmonic strength that maximizes the probability of observing rosettes (see Figure~\ref{fig:2}c).

\subsection{Self-alignment interactions boost the probability of observing  rosettes}
Next, we investigated the effect of self-alignment interactions, modeled via the addition of a term 
to the equation dictating the orientation of the self-propulsion
$\theta_i$, which enforces the alignment between the direction of the self-propulsion velocity and the principal axis of elongation of the cell:
\begin{equation}
 \dot{\theta}_i = - J\sin(\theta_i - \phi_i) + \eta_i
\end{equation}

The angular dynamics is controlled by the interplay of rotational diffusion at rate $D_r$ and alignment at rate $J$, whose inverse $\tau_J=\frac{1}{J}$ is the response time required by the cell to reorient its polarization in the direction of the resultant force exerted by its neighbors~\cite{Paoluzzi2021}.
Figure~\ref{fig:3} shows stationary snapshots of the resulting simulations.
In particular, Figure~\ref{fig:3}a displays representative stationary configurations of populations of cells having different motility in the presence of an external harmonic potential of intensity $k=0.2$  with ($J>0)$ and without ($J=0$) self-alignment interaction.
While increasing the motility tends to impede the formation of rosettes, alignment allows even highly motile cells to organize in rosette-like assemblies.  
In addition, increasing the strength of the alignment interaction boosts the probability of observing transient rosettes at low values of target shape factor (see Figure~\ref{fig:3}b).

\begin{figure*}[]
\centering
\includegraphics[width = 0.8\textwidth]{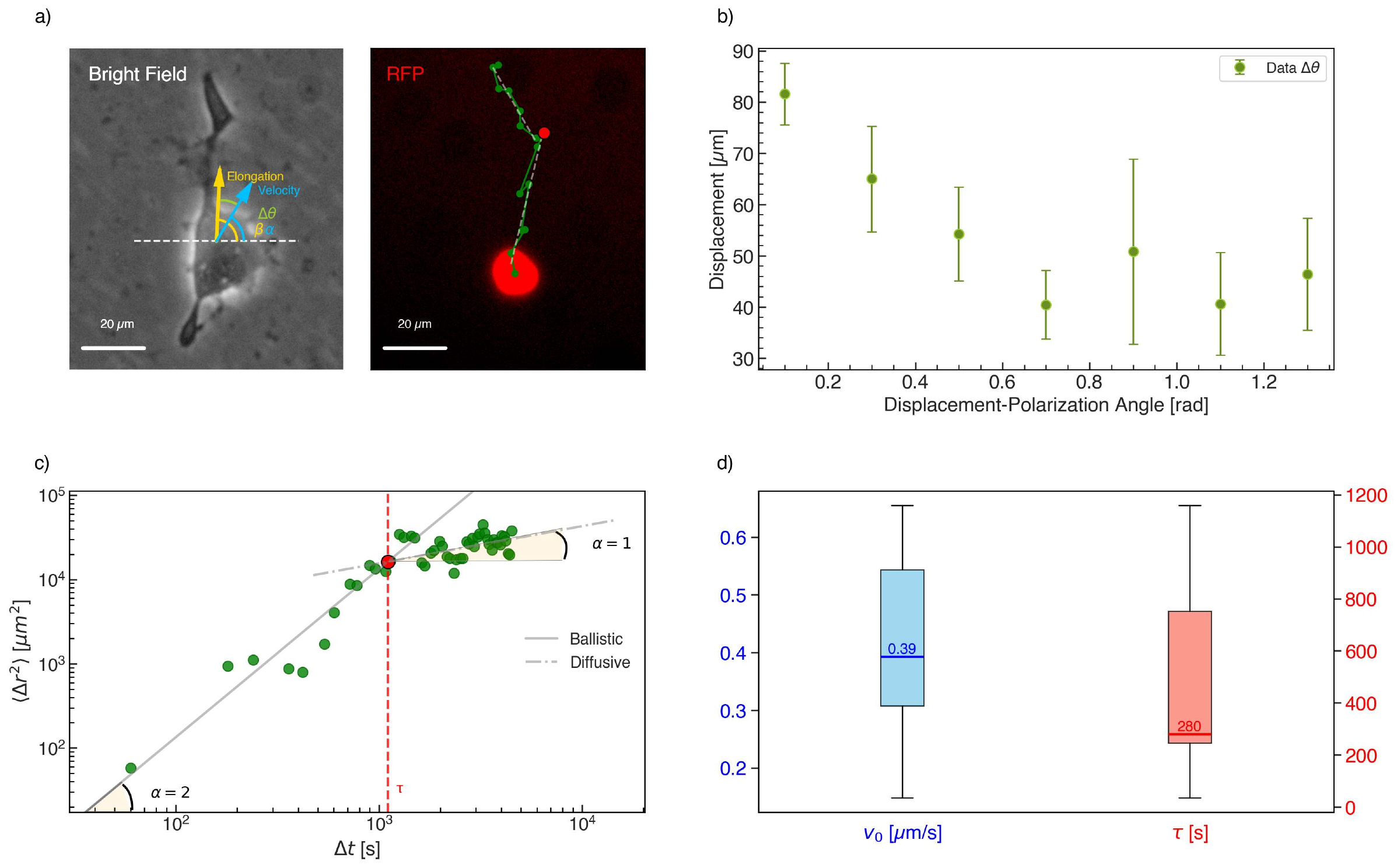}
\caption{\textbf{Measuring dynamical features of neural precursor cells.} \textbf{a)} Bright field (left) and RFP fluorescence (right) microscopy images of iPSCs seeded at low density. Yellow and cyan arrows mark the direction and modulus of the cell self-velocity and polarization, respectively. The green line shows the trajectory of the highlighted cell during the time lapse.  \textbf{b)} Cell displacement as a function of the difference between the displacement and polarization angles.
\textbf{c)}  Mean squared displacement as a function of time for a representative cell dynamics. A sketch of the extracted quantities is reported together with the best fit of Eq.~\ref{eq:msd}. \textbf{d)} Box plot representation of the distribution of measured self velocity (cyan) and persistence times (red).     
} 
\label{fig:6}
\end{figure*}

\begin{figure*}[]
\centering
\includegraphics[width = 0.99\textwidth]{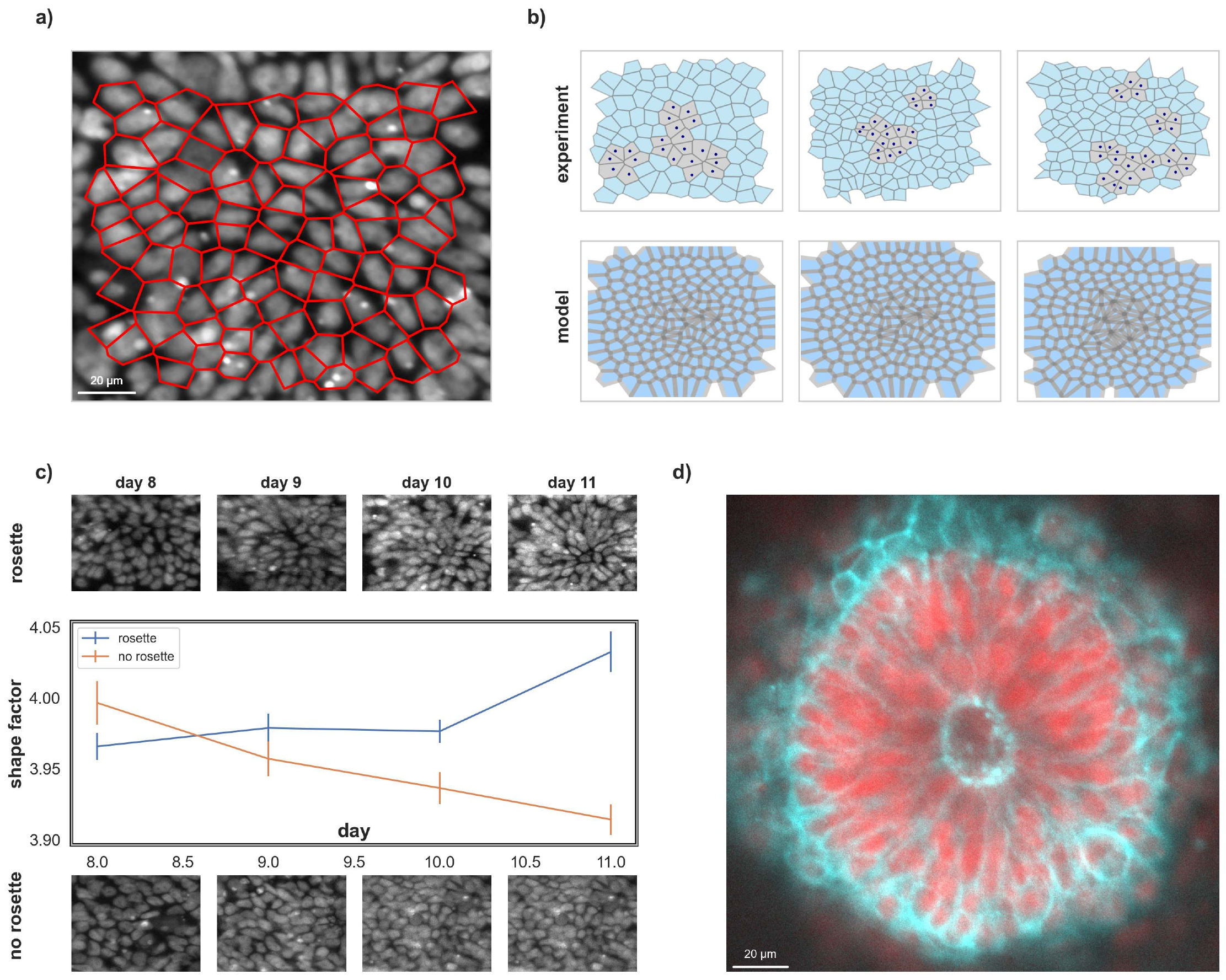}
\caption{\textbf{Neural rosette formation process.} \textbf{a)}  Confocal fluorescence microscopy image of iPSCs expressing the FUS protein fused to a RFP. Red lines represent the best Voronoi tassellation considering cell nuclei as seeds.  \textbf{b)} Comparison between snapshots taken during the temporal evolution of an iPSCs population and a dynamic Voronoi simulation in the presence of a harmonic potential. Transient rosettes are highlighted in gray. \textbf{c)} Average shape factor as a function of time for cell populations that form or do not form rosette structures. Representative snapshots of the considered regions are shown at the top and bottom of the panel for a rosette-forming region and a no-rosette region, respectively.  \textbf{d)} Confocal fluorescence microscopy image of a lumen-centered rosette observed after 11 days from the start of the differentiation procedure on the iPSC cells. Nuclear RFP FUS fluorescence is represented in red, while N-cadherin is in cyan. Scale bar 20 $\mu m$.
} 
\label{fig:7}
\end{figure*}

\subsection{Alignment increases the size of lumen-centered rosettes and influences their polarization}

Transient rosettes can mature in lumen-centered ones~\cite{Harding2014}. We modeled the net effect of the lumen on nearby cells using a Lennard-Jones potential irradiating from one of the Voronoi cell centers. In particular, we imposed that one of the N cells composing the population has zero velocity, vanishingly small target area and perimeter, and exerts a potential energy interaction with the other cells of the population of the form:
\begin{equation}
    E_{i,0} = \epsilon_{i0} \left[ \left( \frac{\sigma_0}{r_{i0}}\right)^{12} - \left(\frac{\sigma_0}{r_{i0}}\right)^6 \right] 
\end{equation}

Figure~\ref{fig:4}a displays a snapshot of a stationary configuration obtained via the Active Voronoi model in the presence of the LJ potential with parameters, $\epsilon_0 = 1.0$ and $\sigma_0= 3$. Cells forming the first, second, and third layers around the central lumen are colored in green, gold, and blue, respectively.  Qualitatively, it can be seen that cells organize into a polarized structure around the lumen. Notably, this behavior is conserved upon varying the lumen size (Figure~\ref{fig:4}b). To get quantitative insight into the role of the model parameters, we characterized the size and organization of the rosette by counting the number of cells that compose the various layers and their relative orientation with respect to the lumen center. In particular, we defined the n-th layer size as $l_n = \sum_i \delta(d_{i0} - n)$, where $d_{i0}$ represents the minimum number of cells one has to go through to reach the center of the Lennard-Jones potential starting from cell $i$. Moreover, we introduced a polarization index for the n-th layer of cells, $\phi_n$, as $\phi_n= \braket{ \Delta_i }_n$, where $\Delta_i$ is defined as the angle between the elongation angle of the cell and the direction of the vector linking the i-th cell center and the center of the lumen. 
Figure~\ref{fig:4}c shows the number of cells in the first cellular layers, $l_1$, as a function of time for different values of the cell target shape factor, $p_0$. After an initial transient time, the number of cells in the layer reaches a stationary state, whose value depends on the target shape factor of the cell population: the higher the target shape factor, the higher the number of cells in the layer. This behavior is evident looking at the stationary values of $l_1$ in Figure~\ref{fig:4}d  as a function of the cell target shape factor, $p_0$, and Lennard-Jones effective radius, $\sigma_0$, for two values of the Lennard-Jones potential strength, $\epsilon_{LJ}$. The number of cells in the rosette increases, not only changing the size of the lumen, as one would expect, but also the strength of the Lennard-Jones potential, i.e., the attraction of the cells toward the lumen. The increase of cells around the lumen reflects in a polarization of the layer with respect to the lumen center (see Figure~\ref{fig:4}e,g).
 As one can see from the snapshots in Figure~\ref{fig:4}f, the size and polarization of the rosette increase according to the values of the cell target shape factor, $p_0$, for a given lumen size. Notably, the number of cells in a given rosette layer and the corresponding polarization of the layer with respect to the lumen are correlated. Intuitively, the more cells in a layer, the more they will tend to assume an elongated form. 
This trend is appreciable by comparing the behavior of the average polarizations of cells in the first cellular layers in Figure~\ref{fig:4}g-h with those in panels c) and d) of the same Figure.
The presence of elongated cells suggests that self-alignment interaction may influence the dynamics of the rosette. To probe this hypothesis, we repeat simulations turning on the self-alignment interaction. 
As one can see both by visual inspection of the configurations in panel i) of Figure~\ref{fig:4} and 
by comparing the distributions of $l_n$ and $\phi_n$ in Figure~\ref{fig:4}l (for a system of 400 cells having a target shape factor of 3.9), the presence of self-alignment interaction significantly increases the number of cells in the first and second layers of the rosette. Indeed,  these layers are composed of $\sim 30$ \% more cells than their counterparts obtained for $J=0$. This difference reduces to $\sim10$ \% for the third layer.

\subsection{ Leader cell-centered rosettes display similar behaviors without a potential term}

So far, we modeled the presence/effect of the lumen through an effective potential energy term, under the assumption that the concentration of actin filaments in the lumen creates an effective steric hindrance of the nearby cells and a net attraction, with cells attempting to reach the lumen region. To test to what extent our findings are dictated by the presence of the potential, we carried out simulations in which the potential is turned off while a leader cell has a different target area, $A^L_0$, with respect to the rest of the population.
Figure~\ref{fig:5}a displays snapshots of stationary configurations obtained via the Active Voronoi model in the presence of a leader cell of target area $A^L_0$. The top row displays snapshots for a population of cells having target shape factor $p_0=3.2$, while the bottom row corresponds to cells with $p_0=3.8$.
Qualitatively, configurations are similar to those observed in the presence of the LJ potential, especially for target areas higher than the target areas of the cells composing the population. Notably, the introduction of a leader cell with a bigger target area provokes the formation of transient rosettes in the close proximity of the leader-cell-centered rosette. The presence of such a cell introduces a defect in the system, which mimics the perturbation produced by the external potential: to accommodate the defect, nearby cells tend to form transient rosette structures.

Evaluating the stationary number of cells and the polarization index for the first cellular layer, we found that rosettes formed around a leader cell manifest a behavior qualitatively similar to those formed in the presence of a LJ potential. In  particular, looking at $l_1$   (green shaded map) and $\phi_1$  (blue shaded map) as a function the cell target shape factor, $p_0$ and leader cell target area, $A_0$, we see that both the size and the polarization of the rosette increases as the population passes from solid-like to liquid-like as observed for the LJ case (see Figure~\ref{fig:5}b-c)
Finally, we probed the effect of self-alignment, retrieving a similar, albeit weak, effect of this interaction on both the size and structure of the rosette. Results are shown in Figure~\ref{fig:5}d-e for the rosette first layers in the absence and presence of self-alignment interaction for different values of cell target shape factor, $p_0$.

\subsection{Neural rosettes are observed in fluidity-preserving 2D cultures}
To test our model predictions, we collected fluorescence microscopy time-lapses during the neural differentiation of iPSCs (see Methods for details). Under the proper differentiation conditions, iPSCs form rosettes after 8-10 days~\cite{Karus_2014}. In the first 4-6 days from the plating, cells are in low-density conditions, thus it is possible to directly measure the dynamical behavior of single cells. Figure~\ref{fig:6}a shows a snapshot of the dynamics of iPSCs tagged with a fluorescent marker (RFP) fused to a nuclear-enriched protein (FUS). Note that fluorescence of the nuclei allows for an accurate tracking of the cellular motion, while low density eases the segmentation of all cells in the field of view (see Methods). Analyzing the dynamics of single cells, we found that they preferentially move along the direction of their elongation. This can be appreciated by looking at
Figure~\ref{fig:6}b, which displays the instantaneous displacement as a function of the angle between the motion direction and the elongation axis. 
Next, we evaluated the mean squared displacement $\langle \Delta r^2\rangle$
as a function of time for all identified trajectories (see Methods). Figure~\ref{fig:6}c shows an example of experimental $\langle \Delta r^2\rangle$ together with the best fit of the expected theoretical behavior $\langle \Delta r^2\rangle_{th}$ in the case of active particles moving with self-propulsion velocity $v_0$ and characterized by a persistence time $\tau$ \cite{RevModPhys.88.045006}
\begin{equation}
\label{eq:msd}
    \langle \Delta r^2\rangle_{th} = 2v_0^2 \left( \frac{t}{\tau} - 1 + e^{-\frac{t}{\tau}} \right) \; .
\end{equation}
According to (\ref{eq:msd}), the cell performs a persistent motion characterized by a ballistic regime in the $\langle \Delta r^2 \rangle$ on short time scales, i.e., for $t \ll \tau$. In the ballistic regime, $\langle \Delta r^2 \rangle$ grows quadratically $\langle \Delta r^2 \rangle \propto v_0^2 t^2$. At larger time scales, i. e., $\tau \gg \tau$, the random walk is characterized by a diffusive behavior so that $\langle \Delta r^2 \rangle$ grows linearly $\langle \Delta r^2 \rangle \propto D \, t$, with $D = v_0^2 \tau /2$ in the case of active motion in two spatial dimensions. As it is easy to check, these two regimes can be obtained by considering a small and large time expansion of (\ref{eq:msd}). $\tau$ sets the crossover time between the two regimes.
For this trajectory, the change from ballistic to diffusive motion is evident from the modification of the slopes at $\tau \sim 250$ s. The results of all tracked cells are reported as box plots in Figure~\ref{fig:6}d. Cells show a self-propulsion of $v_0 = 0.4 \pm 0.1$ $\mu m/s$ and a persistence time of $\tau=260\pm50$ s.  The resulting persistence length is $l_p = v_0 \tau \sim 100 \mu m$. 
In order to compare the behavior of the real data with respect to their \textit{in silico} counterparts, we proceeded to determine the distance scaling factor as follows: we took representative snapshots of the confluent cell population cropping images in LxL square regions (see Figure~\ref{fig:7}a), compute the corresponding Voronoi tesselation (see Methods) and determined the ratio, $f$, between the effective edge length of the region (measured in $\mu m$) and the squared root of the number of cells present in the region ($\sqrt{N_c}$). Note that simulations were performed considering N cells in a square box of $\sqrt{N}$ edge length. This allows us to properly rescale the measured quantities and compare them with simulations. Rescaling of the simulation box equals having cells with an effective length of $\sim \sqrt{A_0} = 1$, while the effective length in the data is $l_c \sim 10 \mu m$.  Thus, $\tilde l_p = f \cdot l_p = 10 l_c$.   

Having fixed the persistence times of the rotational dynamics to $\tau = 50$,  simulations can be compared to real data for values of the self-propulsion velocity around $0.2$. 

Indeed, analyzing the tesselation more in detail, we found that several snapshots display cells in rosette-like arrangements, qualitatively similar to those observed in the simulations performed with the harmonic potential. Figure~\ref{fig:7}b shows some examples of real (top row) and simulated (bottom row) configurations. In particular, it is interesting to note that such transient rosettes are not isolated but form clusters, as we observed in the simulations with cells being in the fluid-like region. Measuring the shape factors of the cells in the snapshots, we found an average shape factor always higher than 3.95, indicating that those configurations are in a fluid-like regime. 
Notably, as shown in Figure~\ref{fig:7}c, we found that cells in regions that do not form rosettes tend to reduce their shape factor in time, while regions on the layer that maintain a shape factor above 3.95 display rosettes that mature in lumen-centered structures (Figure~\ref{fig:7}d). This trend is in accordance with our numerical predictions: the probability of observing transient rosettes is higher for cells in the fluid-like regions, where also the rosette size and polarization are maximum in the lumen-centered case (see Figure~\ref{fig:4}).


\section{Discussions}

In this work, we identified two key ingredients driving the formation of rosettes: local curvature and self-alignment interactions. The role of curvature can be understood by analyzing the configurations associated with the standard Vertex energy in Eq.~\ref{eq:vertex-ene} for a population of cells in a box with periodic boundary conditions, i.e., on a toroid.

In such a flat Euclidean space and with homogeneous parameters, the energetically optimal tiling for the cells is the hexagonal one. 
In fact,  the number of vertices ($V$),  edges ($E$), and cells ($N$) are linked by the Euler's rule as: $V - E + N = \chi$, 
where $\chi$ is the Euler characteristic~\cite{DelBono2024}, equal to 0 in  flat topologies.

If there are no peculiar heterogeneous distributions of line tensions among the cells,   each vertex connects three edges, as T-junctions are optimal to balance the edges' tensions symmetrically.  It follows that $3V = 2E$, and it is straightforward to prove that the average number of edges per cell must be 6, leading to a preference for hexagonal patterns. 


Regular hexagonal tiling actually corresponds to a zero Vertex energy for a value of the target shape index of $s_0 \simeq 3.72$. Different values of shape index introduce topological frustrations in the system, which are compensated by the formation of cells with pentagonal or heptagonal shapes (see polygon distributions in Figure~\ref{fig:1}).  
Each deviation from hexagonal geometry inserts a topological charge that may be compensated via the formation of $>3$-fold vertices. 
In fact, their presence reduces the average number of sides per face.
Cell activity is a promising means to modulate frustration in the cell layer. In particular, by tuning cell self-propulsion velocity, the system is known to display a solid to fluid-like behavior~\cite{Bi2016}.
However, such ingredients are not sufficient to account for the observed presence of rosettes (see Figure~\ref{fig:1}), as in flat surfaces, even four-edge junctions are unstable configurations and rapidly result in T1 transitions,  without special conditions, i.e., peculiar disposition of the line tensions. 

Indeed, from Plateau's laws for soap films,  we know that the angles at a vertex depend on (i) the number of edges meeting, (ii) the tensions in each edge, and (iii) the local curvature of the space.
Thus, higher-fold junctions could, in principle, become favorable for specific anisotropic tension configurations, but also where curvature affects the local angular sum at vertices.   
Specifically, a positive curved space corresponds to a sum of angles exceeding $360^\circ$. As the natural meeting angles between edges change, in this scenario, rosettes become stable because distributing larger angles over more edges reduces mechanical stress.
Moreover, positive curvature induces cells to tighten, which in turn produces cells with a low number of sides (see polygon distributions in Figure~\ref{fig:1} and ~\ref{fig:2}).

Introducing the external harmonic field via  Eq.~\ref{eq:tot_force_harm}, we inserted the simplest positive correction to the curvature. 
To accommodate such curvature, the system tends to introduce topological defects that locally reduce angular excess, forming rosette-like structures (see snapshots in Figure~\ref{fig:2}). Euler's characteristic becomes $\chi > 0$,  modifying the balance of defects:   an excess of positive topological charge is required.


From a physical perspective, defects carry an elastic or energetic cost and act like quasi-particles with topological charge. They interact with each other and with the underlying curvature, minimizing the total energy. This dynamic plays out in soft matter systems, liquid crystals, and 2D materials like graphene~\cite{Wen2013}. 

In ideal, infinite systems with flat geometry, translational symmetry ensures that the structure can be shifted in space without altering its physical properties. 
On curved surfaces,  there is no global translational symmetry, as each point has different geometric properties. 
The introduction of a local curvature and the accompanying defects break this symmetry. 
Defects such as 5- and 7-fold polygons are therefore indicators of the breaking of translational invariance. 

In the present work, we produced such a breaking of the space translational invariance via either a potential energy landscape or a leader cell with a different size with respect to the rest of the population.
In the former case, the external potentials used can mimic the collective effects of apical actin filament networks. Indeed, actin filaments play a fundamental role in shaping the mechanical environment of cells, particularly by forming networks that exert tensile forces across the apical surface of epithelial cell layers. These supra-cellular actin networks can generate coordinated mechanical tension, contributing to tissue morphogenesis and the maintenance of epithelial integrity~\cite{FernandezGonzalez2011}.  In particular, it has been demonstrated how apical actomyosin cables and medial actin structures contribute to mechanical force distribution across cell layers~\cite{Rauzi2010, FernandezGonzalez2011}.
Similarly, the potential can simulate the effect of an actual curved surface where cells are constrained to move, like in embryos and tumor spheroids~\cite{Trichas2012}. 
Note that in these real systems, there is a spontaneous breaking of the symmetry as internal cell mechanisms are responsible for processes like the accumulation of actin fibers or the shape of the embryo. 
Indeed, we show that the presence of a large cell in the layer suffices to insert a topological defect that provokes the formation of transient rosettes.  

Relying on a dynamical model, we were able to probe the stabilizing/destabilizing effects of cellular properties on the formation of rosettes. In particular, we found that cell motility and deformability influence the formation and maturation of rosettes. 
Our simulations highlighted an interplay between topological curvature, self-propulsion velocity, and target shape factor (Figure~\ref{fig:2}b-c). In the presence of fast-moving cells, rosette formation needs both an increase in space curvature and a higher cell deformability. Similarly,  both the size and polarization of lumen-centered structures are favored when the cell population is in a fluid-like state.
Notably, analyzing the behavior of populations of  
iPS-derived neural cells growing on a 2D layer via extensive fluorescent microscopy time lapses, we confirmed our model predictions, as rosettes form in regions characterized by slow-moving cells having a shape factor higher than 3.9, i.e., being in a fluid-like regime. 
Looking at the motion of isolated cells, we observed that their motion is maximum when it aligns with the direction of their elongation (Figure~\ref{fig:6}).  
Self-alignment interactions that produce cell displacements in the same direction along which cells are elongated are the second ingredient we identified that increases the probability of transient rosette formation (see Figure~\ref{fig:3}) and the size of lumen-centered ones (see Figure~\ref{fig:4}). In particular, it allows cells to orderly dispose in multiple layers around the lumen as shown in Figure~\ref{fig:5}.

In conclusion, our work proved that rosettes are natural outcomes of geometric and mechanical responses to curvature, providing an additional framework where the interplay between defects and spatial curvature offers profound insights into how materials and surfaces
are structured, with deep implications across disciplines as diverse as condensed matter physics and cosmology. To test our prediction, we measured the dynamical features of populations of iPSCs and found that neural rosettes are observed when cells populate the region of the parameter space our model predicts to be compatible with their formation.


\section{Materials and Methods}
\label{sec:MatMet}

\subsection{Cell Culture}
Human iPSCs from the WT I line~\cite{Lenzi_2015} were stably transfected with a construct encoding an RFP-tagged version of the nuclear FUS protein~\cite{De_Santis_2019}. iPSCs were induced to differentiate along the neuroectodermal lineage to form neural rosettes as follows. 10.000 cells/cm2 were plated onto glass-bottom dishes (WillCo) coated with Matrigel ( in Nutristem XF/FF medium (Biological Industries) supplemented for 24 hours with 10 µM ROCK inhibitor (Y-27632, Sigma-Aldrich). After 2 days, medium was replaced with KSR medium (DMEM/F12 ; 15\% knockout serum replacement; 1X Glutamax; 1X non-essential amino acids; 1X penicillin-streptomycin; all from Life Technologies). This is considered day 0 of differentiation. From day 4 to day 10, KSR medium was gradually shifted to N2M medium (DMEM/F12; 1X N2 supplement; 1X Glutamax; 1X non-essential amino acids; 1X penicillin-streptomycin; all from Life Technologies). 10 µM SB431542 (Miltenyi Biotec) and 500 nM LDN193189 (Sigma-Aldrich) were added from day 0 to day 7. From day 10 onwards the medium was supplemented with 20 ng/ml BDNF, 20 ng/ml GDNF (PeproTech) and 0.2 mM ascorbic acid were added from day 10 onwards.

\subsection{Immunofluorescence staining}
After 11 days from the beginning of the iPSCs differentiation, cells were fixed in PFA4\% and permeabilized in Triton X-100 0.2\% for 5'. After a blocking stage in BSA 1\% (1h at RT), cells are incubated overnight with the anti-N-cadherin antibody (Ab18203; 5ug/ml) followed by a second stage of hybridization with a Goat anti-Rabbit Secondary Antibody, Alexa Fluor-647 (A-21244).

\subsection{Microscopy experiments}
Measurements were performed via an Olympus microscope with a 20x objective equipped with a V1 Spinning Disk (CrestOptics) and a stage-top CO2 incubator. 
\textit{Time-lapse experiments.} Time-lapse experiments were used for the evaluation of the dynamical features of iPSCs when plated at low density.  
Pictures were taken every 10 min, where in a cycle of 5 acquisitions, 1 acquisition was a bright field picture (BF), and the other 4 were fluorescence acquisitions on the RFP channel (RFP).  
\textit{Time-course experiments.} Time-course experiments were used to follow the formation of rosettes.
Since rosettes emerge sparsely distributed at random locations on the culture plate, we acquired large mosaics from 2 different culture plates.
Specifically, fluorescence images were acquired daily from day 6 through day 11 of the differentiation process. Monitoring a large area of the culture plate, we captured the emergence of more than one rosette at a time.

\subsection{Single cell tracking}

To measure the dynamical features of a single cell from the time-lapses,  nuclei were tracked using a frame-to-frame approach. First, we used the nuclear fluorescence obtained with the reporter RFP protein expressed on the FUS protein to detect single cell nuclei. Second, the nuclei detected in each frame were
connected to the nuclei in the successive frames with links using a nuclei-nuclei distance cost function. Frame-to-frame tracking was implemented using the linear assignment problem (LAP) method. 

\textit{Diffusion Analysis}.
Trajectories obtained with the tracking pipeline described above were used to evaluate the mean square displacements. 
Ballistic and diffusive regimes were identified as the best fit regions for $\langle \Delta r^2(t)\rangle \;\approx\; v_0^2\,t^2  $ and $  \langle \Delta r^2(t)\rangle \;\approx\; A \,t $
where $v_0$ is the self-propulsion velocity.
%
The persistence time $\tau$ has been obtained as the point of intersection between ballistic and diffusive regimes.

\textit{Elongation Analysis.} To compute the elongation axis of the cells,  a directional intensity analysis was implemented on BF images via a custom Python pipeline. 
Nuclear fluorescence was used to determine the centroid of each nucleus, from which the elongation axis was inferred. Evaluation was performed along a discrete set of orientations $\theta \in [0, \pi)$, uniformly spaced. For each direction, a linear segment of $\pm 150$ pixels (300 pixels in total) was sampled symmetrically from the centroid. The average BF intensity was computed along each segment.
In BF imaging, cell boundaries typically produce high-intensity (bright) signals due to light scattering at refractive index gradients from the medium to the cell. Conversely, internal regions and elongated protrusions often present lower contrast and appear darker; the underlying assumption is that directions intersecting prominent cell boundaries will yield higher mean intensity values, whereas directions aligned with elongation, where the cell extends more continuously, will intersect fewer such transitions, resulting in lower average intensities.
Thus, the direction along which the average intensity is minimized is taken as the morphological axis of maximal extension. 
The method provides an axial orientation and is particularly suitable for detecting elongated geometries. 
To complement the intensity‐based orientation, a purely geometric analysis was performed on each cell.  Nuclei centroids were used to seed a watershed segmentation on the corresponding BF images, yielding one binary mask per cell.  Only masks that remained fully contained within the image frame at all time points were retained to avoid boundary artifacts.  
For each mask, we performed a Principal Component Analysis on the covariance of the mask's pixel positions. The first principal component
was used to define the major axis of the ellipse that best fits the cell shape.
From the tracking pipeline, the direction of the  instantaneous displacement, $\alpha_t$ is obtained. The geometric elongation‐axis angle, $\beta$,  was measured as in the same reference.  Finally, the orientation alignment was defined by the difference $|\alpha_t-\beta|$.

\subsection{Voronoi Tesselation and structure factor analysis}
Time-course fluorescence images were cropped to a consistent 400 × 400-pixel ROI for all acquired time points. 
Voronoi tessellations were generated from fluorescence microscopy images of nuclei by manually selecting each nucleus’s center and computing the corresponding Voronoi diagram using custom Python code built on the SciPy library. Regions of interest were manually segmented and saved as label images to isolate neural‐rosette formation within each field of view, and the coordinates of the selected nuclei centers were used to compute Voronoi tesselation.
The average target shape factor has been computed as the ratio of a region’s perimeter to the square root of its area $S_{t}= \frac{p_i}{\sqrt{A_i}}$.
Infinite Voronoi regions were discarded automatically, and only polygons entirely contained within the image boundaries were retained for analysis. \\
\textit{Rosette identification.}
Rosettes were manually identified by analyzing the acquired mosaics. Note that while rosette emergence conditions are being understood, it is difficult to predict where they will actually form. Thus, the detection of rosette emergence in 2D cell cultures required the acquisition of large image mosaics that balance spatial resolution with an extended field of view. This results in large mosaic images that must be manually analysed and are hard to process and visualize. 


\subsection{Simulation details}
We performed Molecular Dynamics simulations of the self-propelled Voronoi model by integrating numerically the stochastic equation for each cell center $\mathbf{r}_i$ using the Euler-Maruyama scheme. As an initial condition, we consider a random uniform distribution of
$N$ cell centers arranged in a square box of side $L=\sqrt{N}$ with periodic boundary conditions. 
Next,  each simulation has been carried out considering $N=400 cells$, for a total time of $10^4$. Equations of motion have been integrated with a time step of $dt=10^{-3}$. Other used dynamical parameters are specified when corresponding results are discussed.
We monitored that the system reached a stationary state by looking at the behavior of the descriptors used to describe rosettes ($P_r$, $l_1$, etc). Typically, stationary configurations are reached after $\sim 10^{4}$ steps.
The algorithm for the analytical computation of forces within the Voronoi model is the one adopted in ~\cite{voronoi2024}, which is the standard fashion for force computation within the Voronoi model of biological tissues \cite{Bi2016,sussman2017cellgpu,barton2017active}.

\section*{Data Availability}
The data that support the findings of this study are available from the corresponding
author upon reasonable request.

\section*{Code Availability}

All codes used to produce the findings of this study are available from the corresponding author upon request.


\subsection*{Competing Interests statement}
The authors declare no competing financial or non-financial interests.

\section*{Acknowledgements}
This research was partially funded by grants from ERC-2019-Synergy Grant (ASTRA, n. 855923); EIC-2022-PathfinderOpen (ivBM-4PAP, n. 101098989); Project `National Center for Gene Therapy and Drugs based on RNA Technology' (CN00000041) financed by NextGeneration EU PNRR MUR—M4C2—Action 1.4—Call `Potenziamento strutture di ricerca e creazione di campioni nazionali di R\&S' (CUP J33C22001130001). M.P. acknowledges funding from the Italian Ministero dell’Università e della Ricerca under the programme PRIN 2022 ("re-ranking of the final lists"), number 2022KWTEB7, cup B53C24006470006.
G.G. acknowledges funding from H2IOSC Project - Humanities and cultural Heritage Italian Open Science Cloud funded by the European Union – NextGenerationEU – NRRP M4C2 - Project code IR0000029 -CUP B63C22000730005.
The authors also thank the support of the staff of the technical office and the microscopy facility of the Italian Institute of Technology.

\bibliographystyle{unsrtnat}
\bibliography{main.bib}

\end{document}